\documentclass[aps,prl,pre,twocolumn,superscriptaddress]{revtex4-2}

\usepackage[pdftex]{graphicx}
\usepackage{lipsum}
\usepackage{SIunits}
\usepackage{xspace}
\usepackage{xcolor}
\usepackage{amsmath}

\newcommand{\etar}{\ensuremath{\eta_\mathrm{r}}\xspace}
\newcommand{\phic}{\ensuremath{\phi^\star}\xspace}
\newcommand{\hs}{\ensuremath{h^\star}\xspace}
\newcommand{\lc}{\ensuremath{\ell_\mathrm{c}}\xspace}
\newcommand{\ls}{\ensuremath{\lambda_\mathrm{c}}\xspace}

\begin{document}

\title{Suspensions of non-Brownian rods: droplet pinch-off, onset of heterogeneity and effective extensional viscosity}


\author{Virgile Thi\'evenaz}
\email[]{virgile.thievenaz@espci.fr}
\affiliation{PMMH, CNRS, ESPCI Paris, Sorbonne Universit\'e, Universit\'e Paris-Cit\'e, F-75005, Paris, France}
\affiliation{Department of Mechanical Engineering, University of California, Santa Barbara, California 93106, USA}
\author{Nathan Vani}
\affiliation{PMMH, CNRS, ESPCI Paris, Sorbonne Universit\'e, Universit\'e Paris-Cit\'e, F-75005, Paris, France}
\affiliation{Department of Mechanical Engineering, University of California, Santa Barbara, California 93106, USA}
\affiliation{Institute of Physics, University of Amsterdam, Science Park 904, Amsterdam, the Netherlands}
\author{Alban Sauret}
\affiliation{Department of Mechanical Engineering, University of California, Santa Barbara, California 93106, USA}
\affiliation{Department of Mechanical Engineering, University of Maryland, College Park, Maryland 20742, USA}
\affiliation{Department of Chemical and Biomolecular Engineering, University of Maryland, 20742 College Park, Maryland, USA}


\date{\today}

\begin{abstract}
    The stretching and pinch-off of a liquid bridge is a simple way to probe when a suspension of particles stops behaving as a continuum. In this study, we consider density-matched suspensions of rigid nylon fibers with aspect ratios (length over diameter) ranging from 2 to 84, and volume fractions $\phi$ spanning the dilute to dense regimes. High-speed imaging of pendant-drop breakup reveals three successive regimes, as previously observed for spherical particles: an equivalent-fluid regime at early times, a dislocation regime corresponding to the separation of the rods, and a final regime controlled by the interstitial liquid once the neck is devoid of rods. The thresholds between these regimes follow the previously proposed scaling for spherical particles, in which the rod length, rather than the rod diameter, is used as the relevant discrete scale. In the equivalent-fluid regime, pinch-off also leads to an effective extensional viscosity that increases with both volume fraction and aspect ratio. This viscosity is not equal to the shear viscosity measured in a parallel-plate rheometer, but both sets of data are well described by Mills' law using a critical volume fraction \phic. Finally, the critical volume fraction \phic decreases monotonically with the aspect ratio and is well captured by an empirical law. These results show that pinch-off is a sensitive probe of continuum breakdown in anisotropic suspensions and that, for rigid rods, the rod length controls the onset of heterogeneous thinning.
\end{abstract}


\maketitle

\section{Introduction}

Capillary pinch-off is one of the simplest flows in which a complex fluid is driven through a hierarchy of shrinking length scales. As a drop detaches from a nozzle, the neck that connects the drop to the needle continuously thins until breakup. For a simple liquid, such as homogeneous and Newtonian ones, this process is described by well-known self-similar regimes. However, in a particulate suspension, the thinning neck eventually becomes sensitive to the discrete nature of the particles. Pinch-off is therefore a direct way to probe the scale below which a suspension can no longer be treated as a homogeneous continuum. This question matters well beyond the canonical pendant-drop geometry, because thinning ligaments and breakup also control spraying, dispensing, coating, and printing processes.\cite{eggers1997,furbank2004drop,furbank2007,bazazi2025emulsion,lohse2022fundamental,sauret2026diw}

For suspensions of non-Brownian spheres, different pinch-off studies have shown that early in the thinning process, the suspension behaves like an equivalent viscous liquid, with an effective viscosity set by the interstitial liquid and the particle volume fraction. Closer to breakup, the dynamics accelerate because the deformation localizes in regions where fewer particles are present, and finally, the dynamics of the neck is governed by the interstitial liquid once particles have been expelled from the most strongly stretched region.\cite{furbank2007,bonnoit2012,mathues2015,chateau2018viscous,thievenaz2021b,thievenaz2022} Recent work has further shown that the onset of this heterogeneous regime can occur when the neck thickness is still much larger than the particle size, which reveals a mesoscopic length scale for the breakdown of a continuum description of suspension rheology.\cite{thievenaz2022} Related heterogeneous breakup has also been reported in other thinning and atomization problems involving dispersed phases.\cite{thievenaz2021viscoelastic,bazazi2025emulsion,chagot2024microfluidic,heshmatzadeh2025pendant}

Rigid fibers are a more difficult case than spheres. They introduce two particle scales, length and diameter, together with rotational degrees of freedom and a much larger set of possible microstructures, from disordered states to flow-aligned ones.\cite{jeffery1922motion,butler2018microstructural,petrie1999rheology} This is not only a model problem: fiber-like particles are common in biological and technological fluids, and even small amounts of them can strongly modify the viscosity, induce shear-thinning or shear-thickening, or even lead to gel-like behavior.\cite{perazzo2017flow}
Depending on the volume fraction,
one speaks of dilute suspensions when particles do not interact with each other;
of semi-dilute suspensions when their interactions are mostly pairwise additive,
and of concentrated or dense suspensions when many-body interactions cannot be neglected.
Suspensions of long fibers can be considered concentrated even for volume fractions of a few percents,
because the reach of hydrodynamic interactions scales with their length.
In shear flow, isolated fibers undergo Jeffery orbits,\cite{jeffery1922motion} while concentrated suspensions often become shear-thinning because the fibers align more strongly at higher shear rates.\cite{mongruel1999shear,bounoua2016shear} The selected microstructure also depends on the forcing history and on the measurement protocol. Shear reversal can disrupt the aligned state,\cite{bounoua2019shear} and different techniques, such as parallel-plate rheometry and falling-ball measurements, can yield markedly different effective viscosities, especially at high volume fractions.\cite{ralambotiana1997viscosity,mongruel1999shear} More generally, for concentrated suspensions of fiber, the effective viscosity must be understood as a flow-dependent quantity, defined within a given geometry and set of boundary conditions.

This strong sensitivity to microstructure is also reflected in theoretical and simulation studies. Analytical descriptions are mainly available in limiting cases, such as dilute or semi-dilute suspensions of slender rods,\cite{batchelor1971stress,shaqfeh1990hydrodynamic,powell1991rheology} and do not directly cover the dense regime considered here. Numerical studies face the same difficulty because fibers bring many degrees of freedom and long-range hydrodynamic interactions, so they are often restricted to relatively small systems or to specific ingredients, such as contact or frictional interactions.\cite{khan2021rheology} Extensional and capillary flows of fiber suspensions, therefore, remain much less documented than shear flows. Capillary bridge and jetting experiments have started to show how alignment, confinement and interactions modify the response in extension,\cite{chateau2021extensional,li2023confinement,subbotin2020multiple} and related coating flows have highlighted the additional role of particle orientation in capillary transport\cite{jeong2023deposition,maddox2024capillary} and drop impact.\cite{rajesh2025impact} Yet the pinch-off from a nozzle of suspensions of non-Brownian fibers remains poorly characterized. In particular, it remains unclear if a suspension of fibers exhibits the same sequence of equivalent-fluid, heterogeneous, and interstitial regimes as spherical particles. In addition, fibers are characterized by their length and their diameter, and we need to determine which particle length scale controls the onset of heterogeneity. 

In this paper, we address these questions by considering the pinch-off from a nozzle of drops of suspensions of nylon fibers in a viscous liquid. We vary the length and diameter of the fibers so that the aspect ratio is in the range $2.1 \leq \lambda \leq 84$, thereby covering short fibers as well as slender ones. We first investigate the transition from an effective Newtonian regime to a non-Newtonian regime. We extend the analysis method we previously applied to suspensions of monodisperse and bidisperse spheres to suspensions of fibers.
\cite{thievenaz2021b,thievenaz2022}
This method enables us to define an effective extensional viscosity in the Newtonian regime,  and to measure the threshold below which this Newtonian description ceases to be valid. Then we compare the extensional viscosity with the shear viscosity measured on the rheometer. We find that these effective viscosities follow a similar trend and that 
notably, the volume fraction dependence is well described by Mills' equation,
\cite{mills1985non}
whose only fitting parameter is the critical volume fraction \phic.
Finally, we show that the variations of \phic\ with the aspect ratio are empirically well
described by a sigmoid function.

\section{Material and Methods}

\begin{figure}[t]
    \centering
    \includegraphics[width=\linewidth]{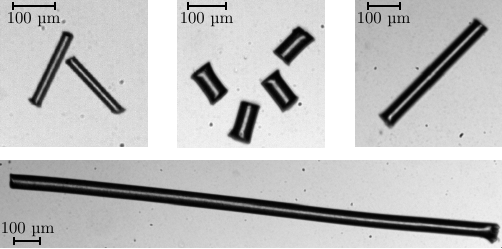}
    \caption{Examples of nylon fibers used in the experiments.
    left: 20$\times$160\micro\meter; 
    center: 50$\times$100\micro\meter;
    right: 50$\times$400\micro\meter;
    bottom: 50$\times$1900\micro\meter.
    }
    \label{fig:rods}
\end{figure}

We use nylon flocking fibers acquired from Cellusuede Products Inc. (Rockford, IL, USA) and
cut by the manufacturer at the desired length.
These rods are roughly cylindrical, with cut marks at both ends (Fig. \ref{fig:rods}).
For simplicity of reading, we refer to each kind of rod using the rounded dimensions given by the manufacturer,
\textit{i.e.} we denote by 20$\times$160 the rods having the diameter 23\micro\meter\
and the length 157\micro\meter.
We measured the actual diameter and length of each type of rod.
We found that the diameter is well controlled, but that the length of the rods has 
a certain distribution around a mean value $L$ with standard deviation $\Delta L$.
Table \ref{tab:rods} summarizes the real dimensions of the rods,
including the coefficient of variation (C.V.) of the rod length, equal to $\Delta L / L$.
The aspect ratio $\lambda = L/D$ is defined as the ratio of length to diameter.
For the current study, we used nine different types of rods 
having two different diameters (23 and 50\micro\meter)
and lengths ranging from 100 to 1900\micro\meter.
The resulting aspect ratio ranges from 2.1 to 84.

\begin{table}[t]
    \centering
    \begin{tabular}{|l|r|r|r|r|r|}
        Name & $D$ (\micro\meter) & $L$ (\micro\meter) & $\Delta L$ (\micro\meter) & C.V. & $\lambda$ \\
        20$\times$160  & 23 & 157  & 10  & 6.3\%  & 6.8   \\
        20$\times$400  & 23 & 423  & 130 & 30.4\% & 18.4  \\
        20$\times$760  & 23 & 691  & 110 & 15.9\% & 30    \\
        20$\times$1900 & 23 & 1929 & 49  & 2.5\%  & 84    \\
        50$\times$100  & 50 & 105  & 8   & 7.6\%  & 2.1   \\
        50$\times$200  & 50 & 221  & 11  & 5\%    & 4.4   \\
        50$\times$400  & 50 & 453  & 15  & 3.4\%  & 9     \\
        50$\times$1000 & 50 & 1078 & 43  & 4\%    & 21.6  \\
        50$\times$1900 & 50 & 1971 & 139 & 7\%    & 39.4  \\
    \end{tabular}
    \caption{Dimensions of the nylon rods, measured using the microscope;
        diameter $D$, average length $L$, standard deviation of the $\Delta L$, 
        coefficient of variation $\Delta L / L$ and aspect ratio $\lambda = L/D$.    }
    \label{tab:rods}
\end{table}

\begin{figure*}[t]
    \centering
    \includegraphics[width=\linewidth]{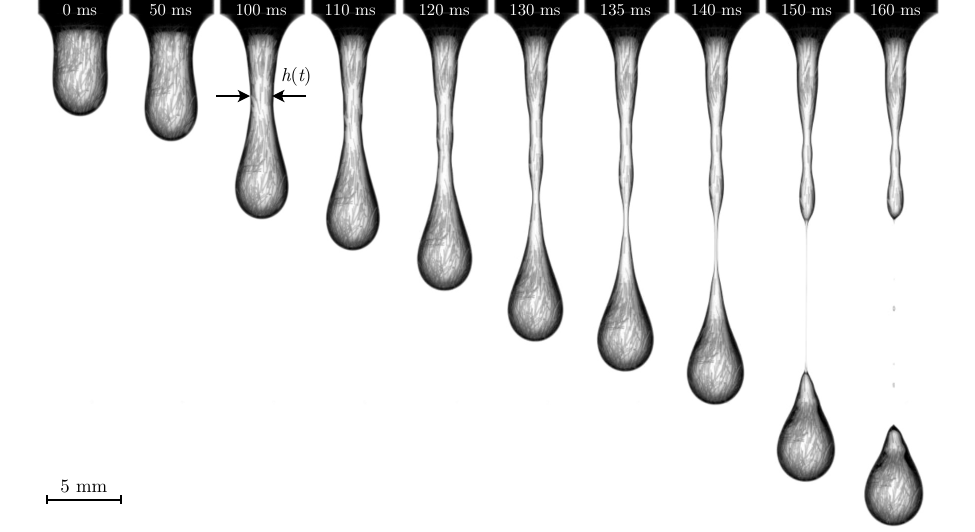}
\caption{Stages of the detachment of a droplet of a suspension of rods.
        Initially, the rods are uniformly distributed; the suspension behaves as a homogeneous Newtonian liquid.
        At some point (about \unit{110}\milli\second), the fluctuations become noticeable.
        Regions of low volume fraction appear; most of the deformation is borne by these regions.
        This is the dislocation of the suspension.
        Eventually (after \unit{135}\milli\second), there is no solid in the region that actually deforms.
        The dynamics revert to that of a Newtonian liquid, the interstitial liquid.
        The corresponding video is available as supplementary material.
    }
    \label{fig:pics}
\end{figure*}

The suspending liquid is a mixture of water (46.2\% wt.),
poly-(ethylene glycol ran propylene glycol) monobutyl ether (45\% wt., molar weight of about 
\unit{3900}\gram\per\mole, Sigma-Aldrich ref. 438189)
and Zinc Chloride (8.8\% wt., Sigma Aldrich ref. 208086).
%
This mixture is a Newtonian fluid with a viscosity $\eta_0 = \unit{270}\milli\pascal\cdot\second$, a density $\rho=\unit{1120}\kilo\gram\per\cubic\meter$ matching that of the nylon rods, and a surface tension $\gamma \simeq \unit{58}{\milli\newton\per\meter}$, which yields a capillary length $\lc \simeq \unit{2.3}{\milli\meter}$. Using density-matched rods suppresses sedimentation and buoyancy-driven migration within the suspension. Inertia remains small over the present conditions because of the large viscosity of the suspending liquid.

%

The rods can be considered rigid throughout the flow.
To justify this, we consider the bending of an elastic rod subject to viscous forces,  
bending being the softest deformation mode for slender bodies. 
The viscous force is $F_\mathrm{v} \sim \eta_0LU/\ln{\lambda}$,
with a typical velocity $U \sim \dot{\varepsilon} L$,
where $\dot{\varepsilon}$ is the macroscopic rate of strain~\cite{cox1970motion}. 
This is to be compared to the force $F_\mathrm{b} \sim  ED^4/L^2$ required to bend the rod
with a radius of curvature of the order of its length.
The elastic modulus $E$ of nylon is about $3\,\mathrm{GPa}$,
and the rate of strain in our experiments is of order $100\,\mathrm{s}^{-1}$ at most.
The ratio of these forces, which scales as $\eta_0\dot{\varepsilon}\lambda^4/(E\ln{\lambda})$,
is at most of the order of 0.1 for the largest aspect ratio $\lambda = 84$ considered here.

Suspensions were prepared by adding rods to the liquid in \unit{20}\milli\liter\ glass vials.
A roller mixer was used to gently mix the rods with the liquid.
A vortex mixer was also used to accelerate the mixing in the cases 
of high aspect ratios at high concentrations.
In all cases, the vials of suspensions were left on the roller mixer for a few tens of minutes 
so that the rod concentration in the suspension was uniform at the start of the experiment.
The shear rheology of the suspensions was characterized using a parallel-plate rheometer 
(Anton Paar MCR 302).
The gap between the plates was set to \unit{1}\milli\meter\ for all measurements. We should emphasize that this gap is large relative to the rod diameter across all suspensions, but it becomes comparable to or smaller than the rod length for the longest rods. The rheometer measurements should therefore be interpreted as geometry-dependent effective viscosities, especially at the largest aspect ratios, where confinement and wall-induced alignment may affect the microstructure.

The pinch-off apparatus consists of a syringe, 
mounted with a stainless steel capillary of outer diameter $h_0=\unit{5.5}\milli\meter$,
and held by a clamp in front of a light panel.
%
Experiments consist of extruding a small amount of suspension through the syringe to create
a pendant drop.
The drop destabilizes when its weight exceeds the capillary force holding it at the nozzle.
As the drop falls, the neck that binds it to the nozzle stretches until it breaks (Fig. \ref{fig:pics}).
The detachment of the drop is recorded using a high-speed camera (Phantom VEO 710)
and a macro lens (Nikon Micro-Nikkor AI-s 200mm f/4).
Each experiment with a given kind of rod and a given volume fraction was repeated five times and
experiments without rods were repeated ten times.

\section{Thinning dynamics}

\begin{figure*}[t]
    \centering
    \includegraphics[width=.99\linewidth]{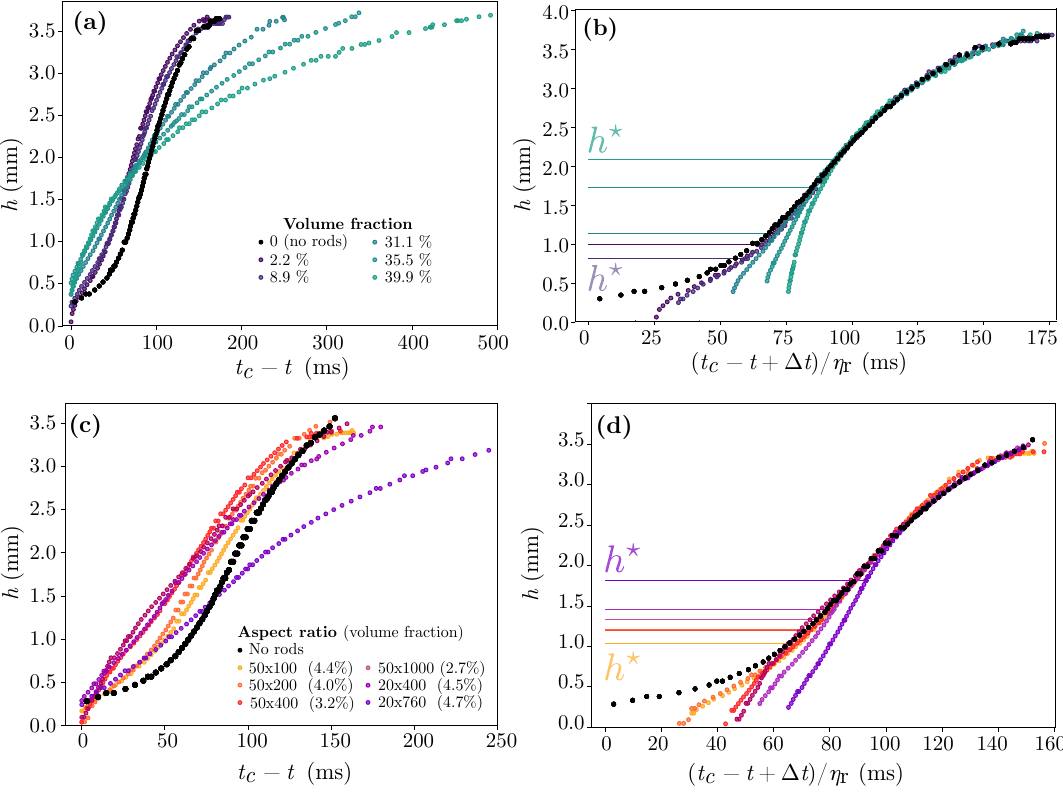}
    \caption{Thinning dynamics of suspensions of rods. 
        (a) represents suspensions of 50$\times$100\micro\meter\ rods with various volume fractions,
        (b) is the best rescaling of (a) onto the Newtonian regime for the early time dynamics,
        using a linear time mapping.
        (c) represents suspensions with rods of various dimensions whose volume fraction is around 4\%,
        (d) is the best rescaling of (c) onto the Newtonian regime for the early time dynamics,
        using a linear time mapping.
        Time goes from right to left.
    }
    \label{fig:dyn}
\end{figure*}

A typical pinch-off experiment with a medium aspect ratio (50$\times$1000\micro\meter)
is shown in Fig. \ref{fig:pics}.
In the first image, the suspension looks relatively homogeneous.
The rods are apparently aligned, likely due to nozzle extrusion. 
As the drop falls and the neck thins down, one begins to observe inhomogeneities in the rod distribution
(see at 110\milli\second).
The neck becomes a region of lower volume fraction, so the local viscosity diminishes.
The suspension becomes easier to deform around this region.
Eventually, there are no rods left in the neck (at about 135\milli\second);
what follows is the classic thinning of a Newtonian viscous liquid \cite{shi1994cascade}.
Despite the anisotropic shape of the rods, we find that the thinning dynamics of the present suspensions
are quite similar to those of suspensions of non-Brownian spheres in a viscous liquid.\cite{thievenaz2021b,thievenaz2022}

We study the thinning of the suspensions by measuring the thickness of the neck at its thinnest point,
$h(t)$ (Fig. \ref{fig:dyn}).
The variations of $h$ in time are related to the local stress at the neck,
where the flow is purely extensional.
Figure \ref{fig:dyn}(a) reports the thinning dynamics $h(t)$ for
suspensions of the same 50$\times$100 rods at different volume fractions.
Time flows from right to left, the origin of time $t=t_\mathrm{c}$ corresponding to the breakup.
The black curve represents the thinning of the pure liquid without rods;
for this curve, the origin of time is arbitrary.
The first general observation is that starting from the same thickness of about \unit{3.6}\milli\meter,
a suspension with a small amount of rods (\textit{e.g} 2.2\%) will thin down and break faster
than the suspending liquid alone.
The difference is evident in the last instants of the pinching: 
the thinning of the suspensions accelerates, whereas for the pure liquid, it does not.
At higher concentrations, the thinning takes longer because the suspension is more viscous,
but at the last stage, it accelerates dramatically.
This acceleration is known to occur in the case of spherical particles.\cite{bonnoit2012}
It corresponds to the decrease of particle volume fraction around the neck \cite{thievenaz2022}.

Bonnoit \textit{et al.} have shown that, with regard to the detachment of a drop,
a suspension of non-Brownian spheres in a viscous liquid behaves 
like a more viscous yet Newtonian liquid, down to a certain scale.\cite{bonnoit2012}
In our previous study\cite{thievenaz2022},
we have shown that comparing the thinning of the suspensions to that 
of the pure suspending liquid leads to a good collapse of $h(t)$ down to a certain scale.


Given the size of the neck, the viscosity, and the typical duration of an experiment,
the thinning results mainly from the balance between gravity (the weight of the drop pulling on the neck)
and the viscosity of the suspension that resists the deformation.
The typical time scale of a given experiment is therefore $\eta / \rho g h_0$.
Since the time scale is linear in the viscosity,
the ratio of the time scales of two experiments with two different liquids
is equal to the ratio of the viscosities of these two liquids.
In particular, the ratio of the time scales $\tau$ and $\tau_0$ respectively corresponding to the pinch-off
of a drop of suspension and to the pinch-off of a drop of pure liquid
should be equal to the relative viscosity
of the suspension: $\tau/\tau_0 \sim \eta/\eta_0 = \etar$.

Inertia at the neck can be neglected during most of the experiment,
although it plays an important role in the very last stages of the thinning of the viscous thread
(corresponding to $t = 150\milli\second$ in Fig. \ref{fig:pics}).\cite{eggers1993}
Taking the time scale of the whole experiment $\tau$ of the order of \unit{100}\milli\second,
the Reynolds number $Re = \frac{\rho h_0^2}{\tau \eta_0}$ is at most equal to one 
in the case of the interstitial liquid,
and much smaller for dense suspensions.

Figure \ref{fig:dyn}(b) shows the same thinning dynamics $h(t)$ after shifting the time axis by some duration $\Delta t$ and stretching it by some factor \etar.
The values of $\Delta t$ and \etar are fitted for each suspension (a given colored curve)
so that the dynamics overlap with the pure-liquid case (the black curve) over the widest possible range.
Using such a linear mapping, we can always match the two dynamics down to a certain scale \hs.
Therefore, for each type of rod and each volume fraction,
we can find the value of the relative viscosity \etar\ and the critical thickness \hs.
Since each experiment is repeated several times, we average the values of
\etar\ and \hs\ over the repeated experiments.

The aspect ratio has an effect comparable to the volume fraction.
Figure \ref{fig:dyn}(c) reports the thinning dynamics for suspensions of different rods with 
similar volume fractions (around 4\%).
The longer the rods are, the slower the overall dynamics is, but the stronger the final acceleration effect.
The corresponding rescaled dynamics are shown in Fig. \ref{fig:dyn}(d).

\section{Threshold to the dislocation regime}

\begin{figure}[t]
    \centering
    \includegraphics[width=\linewidth]{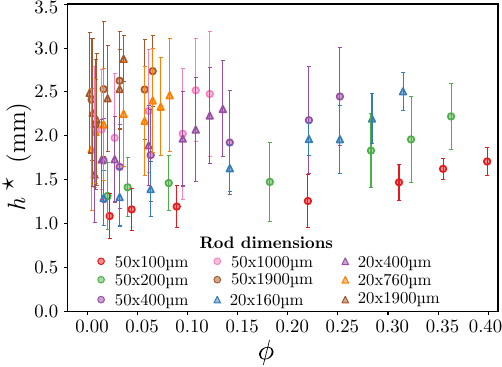}
    \caption{Variations of the threshold \hs with the dimensions of the rods and their volume fraction.
        \hs is the thickness such that for $h>\hs$, the suspension behaves as an effective Newtonian liquid,
        and for $h<\hs$ it does not.
        \hs increases slightly with the volume fraction, and strongly with the aspect ratio.}
    \label{fig:hs}
\end{figure}

\begin{figure}[t]
    \centering
    \includegraphics[width=\linewidth]{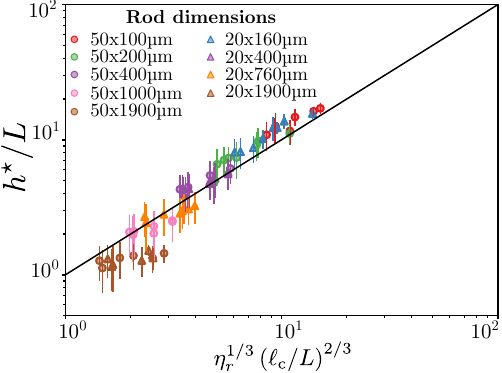}
    \caption{Validation of the scaling law for \hs
        \cite{thievenaz2022} (Eq. \ref{eq:scaling}) for suspensions of rods.
        In non-dimensional form, 
        the data for all kinds of rods and volume fractions (Fig. \ref{fig:hs})
        collapse onto the master curve predicted by Eq. \ref{eq:scaling}.
        The limitations occur when \hs is of the order of the rod length ($\hs \simeq L$); 
        this happens for the longest rods (brown symbols);
        in this case the suspension could never be considered homogeneous.
    }
    \label{fig:scaling}
\end{figure}

\begin{figure*}[p]
    \centering
    \includegraphics[width=.8\linewidth]{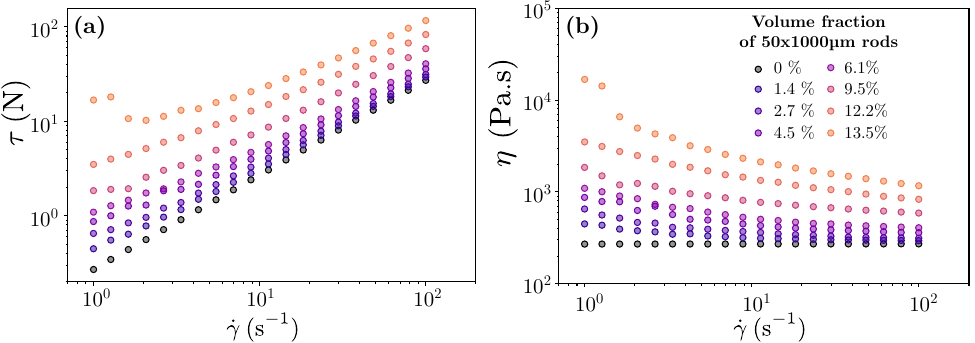}
    \caption{Rheological curves for suspensions of 50$\times$1000\micro\meter\ rods
        with volume fractions ranging from 0 to 13.5\%.
        (a) shear stress $\tau$ \emph{vs.} shear rate $\dot\gamma$,
        (b) viscosity $\eta$ \emph{vs.} shear rate.
        Suspensions of other types of rods show similar behavior.
    }
    \label{fig:rheo}
\end{figure*}
\begin{figure*}[p]
    \centering
    \includegraphics[width=\linewidth]{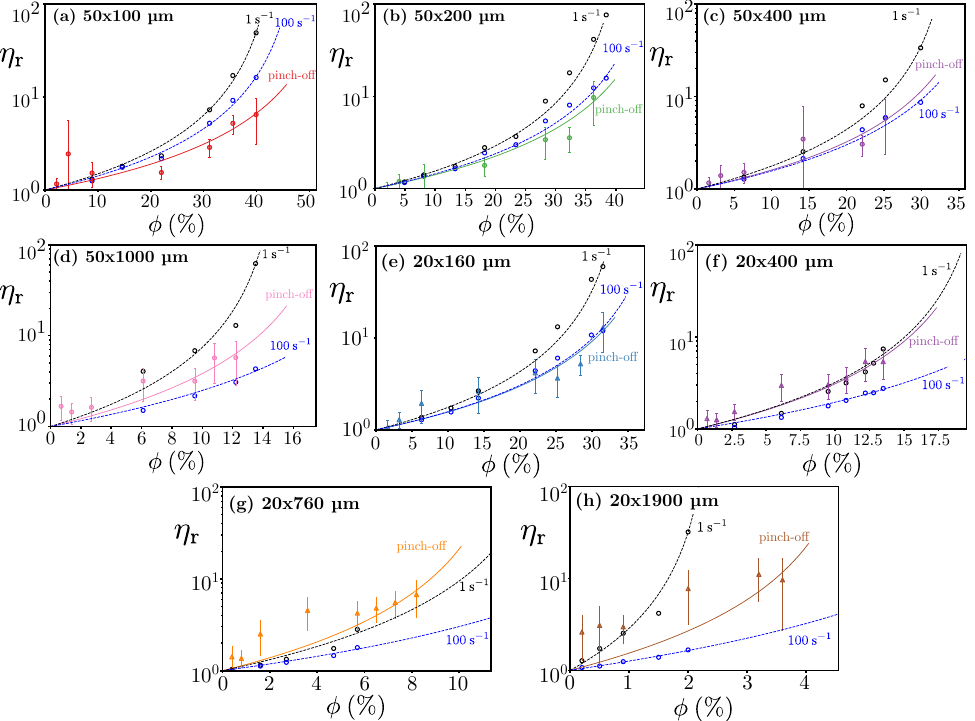}
    \caption{Comparison between the relative viscosity measured in the rheometer
    at $\dot\gamma=$\unit{1}\second\reciprocal\  (black circles) and
    at $\dot\gamma=$\unit{100}\second\reciprocal\  (blue circles),
    to the effective viscosity measured by the pinch-off experiments (other coloured symbols).
    Each line corresponds to the best fit of the like-coloured points by Mills' law (Eq. \ref{eq:mills}). 
    The error bars represent the standard deviation in the pinch-off measurements.}
    \label{fig:rheocomp}
\end{figure*}

The critical thickness of the neck \hs is the threshold under which the suspension behaves 
in a significantly different way from the pure liquid.
This threshold depends on the volume fraction and the dimensions of the rods,
as shown in Fig. \ref{fig:hs}.
We notice that even for suspensions of short rods (50$\times$100, red circles),
\hs\ is of the order of ten times the rod length.
In fact, the threshold to heterogeneous dynamics is not the scale of the particles but
an intermediate scale between the particles and the drop.\cite{thievenaz2022}
For long rods (20$\times$1900 and 50$\times$1900\micro\meter),
\hs\ is larger in absolute terms but smaller when normalized by the rod length.
For a given kind of rod, \hs increases with the volume fraction.

The variations of \hs have been described and interpreted 
for suspensions of spherical particles;\cite{thievenaz2022}
here, we apply the same method to suspensions of rods.
\hs describes the crossover between two flow regimes.
In the first regime ($h>\hs$), the particle volume fraction is uniform around the neck;
the suspension can be described as a homogeneous, Newtonian liquid with an effective viscosity.
This is the effective Newtonian regime.
In the second regime ($h<\hs$), the particle volume fraction fluctuates significantly around the neck;
the stretching is, therefore, concentrated in the region with the fewest particles,
that is the least viscous region of the suspension.
This second regime is the dislocation of the suspension: particles are drawn apart from each other.
After the dislocation, there is a third regime when the neck is devoid of particles altogether;
this is simply the thinning of a viscous liquid,
and there is nothing original in the present configuration.

Each regime corresponds to a certain rate of energy dissipation in the neck,
depending on how the strain is localized.
In the effective Newtonian regime, the strain is uniform in the neck.
In the dislocation regime, the strain is concentrated in a region of the neck smaller than $h$,
whose size corresponds to the length scale of the fluctuation of volume fraction.
The rate of energy dissipation for each regime can be evaluated.\cite{thievenaz2022}
Considering that the transition occurs when it becomes favorable (less dissipative) to dislocate
rather than to deform the neck uniformly, we obtain the scaling law for \hs:
\begin{equation}
    \frac{\hs}{L} \sim \etar^{1/3} \left( \frac{\lc}{L} \right)^{2/3},
    \label{eq:scaling}
\end{equation}
where $\lc = 2.3\milli\meter$ is the capillary length.

Figure \ref{fig:scaling} shows the same data as Figure \ref{fig:hs}, 
plotted in the dimensionless form suggested by Eq. \ref{eq:scaling}.
The collapse of the data points together is excellent. 
The black line corresponds to a perfect match.
As in our previous work on spherical particles,\cite{thievenaz2022} 
we observe that the agreement is less good when $\hs \simeq L$,
because the rods are too long for the suspensions to be homogeneous.

Eq. \ref{eq:scaling} was derived for a suspension of spheres
whose typical length scale is their diameter.\cite{thievenaz2022}
We find that this scaling law is valid for suspensions of rods if one takes their length $L$
as the typical length scale.
It seems that the diameter of the rods has little or no effect on \hs.
This is likely due to the fact that the transport of momentum by a rod in the flow is much more efficient
lengthwise than crosswise.
In the suspension, there are two mechanisms to transmit momentum,
either through the viscous interstitial fluid, or through the rods.
The latter mechanism is not dissipative and thus much more efficient. 
For spherical particles, both mechanisms are isotropic,
but for rods, there is a prefered direction, lengthwise.

\section{Effective viscosity}

The second quantity of interest that the pinch-off experiments enable us to measure 
is the effective viscosity \etar\ in the effective Newtonian regime.
For each suspension, we compare this effective extensional viscosity
to the shear viscosity measured in a parallel-plate rheometer.
 
Figure \ref{fig:rheo} shows typical rheological curves 
((a) stress \emph{vs.} rate of shear, (b) viscosity \emph{vs.} rate of shear)
obtained for rods with a given aspect ratio (50$\times$1000\micro\meter) for various volume fractions.
We find that the suspensions are quasi-Newtonian at low concentrations 
and shear-thinning at higher concentrations.
Similar curves are obtained for all aspect ratios.

In Fig. \ref{fig:rheocomp}, we compare the effective viscosity extracted from the equivalent-fluid regime of pinch-off with the apparent shear viscosity measured in the parallel-plate rheometer. The two quantities are of the same order of magnitude and display the same overall trends with volume fraction and the aspect ratio. They should not, however, be expected to coincide numerically. Pinch-off probes an extensional free-surface flow with a time-dependent strain-rate history and evolving rod orientation, whereas rheometry probes a wall-bounded shear flow. The pinch-off measurements therefore provide an effective extensional viscosity in the sense of a geometry- and flow-dependent stress-to-strain-rate ratio, rather than a direct counterpart of $\eta(\dot\gamma)$ measured at one imposed shear rate. We observe here that depending on the rod dimensions, the same pinch-off configurations lead to effective viscosities smaller or larger than the shear viscosity at shear rates
\unit{1}\second\reciprocal\ \ and \unit{100}\second\reciprocal\, without a well-defined trend.
Those values of shear rates are similar to the typical minimum and maximal values of the 
extension rate in the pinch-off dynamics that are depicted in Figure \ref{fig:dyn}.
The error bars in Fig. \ref{fig:rheocomp} represent the standard deviation across repeated pinch-off experiments for a given suspension.
This deviation is smaller than the difference between the shear viscosities measured at the two shear rates.

We use Mills' equation \cite{mills1985non} to fit the volume-fraction dependence of the viscosity: 
\begin{equation}
    \etar = \frac{1-\phi}{\left( 1 - \phi/\phic\right)^2}.
    \label{eq:mills}
\end{equation}
This equation resembles the Maron-Pierce equation $\etar = \left( 1 - \phi/\phic\right)^{-2}$,
notably used by Bounoua \emph{et al.} to fit their rheology curves for suspensions of rods.
\cite{bounoua2019shear,mueller2010rheology}
Mills' equation is based on the simple and explicit argument that the energy dissipation
at the macroscopic scale must equal the dissipation at the microscopic scale. 
The critical volume fraction \phic is a geometric parameter that describes 
the local environment of the particles.
We use Mills' equation here because it includes the factor $(1-\phi)$ and can be related more directly to the local deformations of the suspending fluid in dense suspensions.\cite{thievenaz2025caging}
Nevertheless, the present data using either Mills' equation or the Maron-Pierce equation gives an equally
good agreement, just slightly different values for \phic.
We obtain a good agreement between experiments and Eq. \ref{eq:mills} for all rheometer measurements,
and for most pinch-off experiments.
Pinch-off experiments with the longest rods (Fig. \ref{fig:rheocomp}h) do not fit well,
probably for the same reason that they do not match Eq. \ref{eq:scaling} in Fig. \ref{fig:scaling}:
the rods are too long for the suspension to be considered homogeneous in the first place.

\section{Critical volume fraction \phic}

\begin{figure*}[t]
    \centering
    \includegraphics[width=\linewidth]{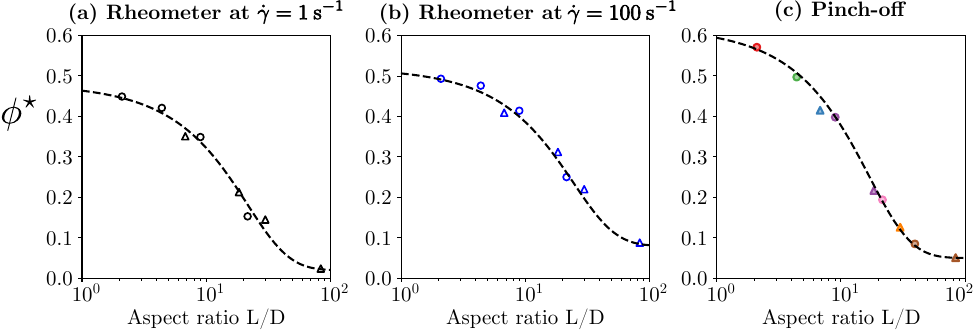}
    \caption{Critical volume fraction \phic\ vs. aspect ratio of the rods.
        \phic is obtained by fitting the viscosity vs. volume fraction curves of Fig. \ref{fig:rheocomp},
        for the three different experiments :
        (a) parallel-plate rheometer at $\dot\gamma =$\unit{1}\second\reciprocal\:;
        (b) parallel-plate rheometer at $\dot\gamma =$\unit{100}\second\reciprocal\:;
        (c) droplet pinch-off.
    Dashed lines represent the best fit by the proposed empirical sigmoid law (Eq. \ref{eq:sigm})}
    \label{fig:phic}
\end{figure*}

An interesting trend can be found when considering the value of \phic\ obtained by fitting the
different viscosity curves.
Figure \ref{fig:phic} shows how \phic varies with the aspect ratio of the rods.
Each subfigure represents a given experimental protocol:
parallel-plate rheometer at shear rate \unit{1}\second\reciprocal\ (a),
rheometer at \unit{100}\second\reciprocal\ (b), and pinch-off experiments (c).

First, for a given experimental protocol, all the fitted values of $\phi^\star$ lie on a single, well-defined curve when plotted against aspect ratio. This means that for the experiments presented here, we do not observe an independent effect of rod length and rod diameter beyond their combination through $\lambda = L/D$. Because the fitted value of $\phi^\star$ depends on the measurement protocol, we do not interpret it here as a unique material constant. Instead, $\phi^\star$ is best viewed as an effective value that depends on both particle geometry and the flow-induced microstructure selected by the experimental configuration.
We find that the master curve is well described by the empirical law
\begin{equation}
    \phi^\star (\lambda) = \phi^\star_\infty + \frac{2\left(\phi^\star_0 - \phi^\star_\infty\right)}{1 + e^{\lambda/\ls}}.
    \label{eq:sigm}
\end{equation}
This form captures the monotonic decrease of $\phi^\star$ with aspect ratio on the logarithmic axis used in Fig. \ref{fig:phic}. It introduces three fitting parameters: the two asymptotic values $\phi^\star(0)=\phi^\star_0$ and $\phi^\star(+\infty)=\phi^\star_\infty$, and a characteristic aspect-ratio scale \ls governing the crossover.
The best-fit values of the parameters are summarized in Table \ref{tab:sigm}.
The set of equations \ref{eq:mills} and \ref{eq:sigm} provides a good empirical model for the
viscosity of suspension of rods as a function of the aspect ratio and volume fraction.

\begin{table}
    \centering
    \begin{tabular}{|l|c|c|c|}
        Method & $\phi^\star_0$ & $\phi^\star_\infty$ & \ls  \\
        Rheometer -- \unit{1}\second\reciprocal & 0.46 & 0.02 & 14\\
        Rheometer -- \unit{100}\second\reciprocal & 0.44 & 0.08 & 16\\
        Pinch-off & 0.57 & 0.05 & 11 \\
    \end{tabular}
    \caption{Fitting parameters for Eq. \ref{eq:sigm} and the data of Fig. \ref{fig:phic}.}
    \label{tab:sigm}
\end{table}

It is notable that two \emph{a priori} very different experiments, parallel-plate shear rheometer and droplet pinch-off, lead to the same qualitative dependence of $\phi^\star$ on aspect ratio. The difference lies in the fitted parameters of the sigmoid law, which suggests that the same fibers organize differently under different boundary conditions and flow types. In that sense, $\phi^\star$ appears to quantify a flow-dependent maximum packing fraction rather than a universal packing limit. Direct measurements of rod orientation and local structure near the thinning neck would help test this interpretation, but such measurements are beyond the scope of the present study.

\section*{Conclusions}

In this paper, we have investigated the pinch-off of drops of suspensions of rigid fibers and shown that, despite the anisotropy of the particles, the thinning dynamics follows the same broad sequence as for suspensions of spheres and can be described by the same methods.
The thinning of the suspension neck initially follows an effective Newtonian regime.
An effective viscosity can be defined from the time scale of this regime.
It follows the same trend as the shear viscosity measured in a parallel-plate rheometer,
although its value is markedly different.

Under a certain threshold \hs, the behavior of the suspensions ceases to be Newtonian.
The suspension dislocates, \emph{i.e.} the rods move away from each other inside the neck;
the local decrease in volume fraction leads to a decrease in the viscosity,
hence the acceleration of the detachment of the drop.
We find that \hs follows the same scaling law (Eq. \ref{eq:scaling}) 
that was derived and confirmed for suspensions of spheres in our previous work. \cite{thievenaz2022}
The relevant length scale to describe the rods in this regime is their length.
Figure \ref{fig:scaling} shows that rods with the same length but different diameters 
find the same location on the scaling law.
However, the rod diameter has an indirect effect on \hs through the effective viscosity, 
that depends on the rod aspect ratio.

The rheological properties of suspensions of rigid rods are generally discussed
relative to the orientation of the microstructure.\cite{butler2018microstructural}
Depending on the volume fraction, the suspension may be dilute (no interactions between rods),
semi-dilute (weak pair interactions), concentrated isotropic (many-body interactions, no ordering),
or liquid crystalline (nematic order: rods aligned in one direction with no other symmetry).
Our results suggest that, for a given class of microstructure (isotropic or nematic), the volume-fraction dependence can be parameterized to first order by the single parameter \phic.
Indeed, both the pinch-off experiment -- which is likely to favor alignment near the neck --
and the parallel-plate rheometry experiment -- which selects a different microstructure --
are reasonably well described by Mills' equation.
In the future, it might prove useful to compare the values of \phic\
with microstructural order parameters.

\section*{Author contributions}
VT and AS designed the research. NV conducted the droplet pinch-off experiments. 
VT conducted the rheology measurements, analyzed the data, and wrote the manuscript.

\section*{Conflicts of interest}
There are no conflicts to declare.

\section*{Videos}
Nine videos illustrating the pinch-off of different suspensions, are provided to supplement Figure 2.
All videos were recorderd at 2000 frames per second, and are played at 20 frames per second (slowed down one hundred times).
For scale reference, the nozzle is \unit{5.5}\milli\meter\ wide.

\section*{Data availability}
The data related to this study is available from the arXiv repository, distributed among the following files:\\
-- \emph{dislocation.csv} contains the data relative to the dislocation regime ($\Delta t$, \etar and \hs), shown in Figures 3 to 5;\\
-- \emph{phi\_star\_vs\_aspect\_ratio.csv} contains the data displayed in Figure 8;\\
-- the \emph{rheology/} folder contains the rheometry measurements performed using the parallel-plate rheometer,
a portion of which is shown in Figure 6 and 7.

\section*{Acknowledgments}
This material is based upon work supported by the National Science Foundation
under NSF CAREER Program Award CBET Grant No. 1944844 and by the American Chemical Society Petroleum Research Fund PRF No. 69673-ND9.

\bibliography{rods.bib}

\end{document}